\documentclass{INTERSPEECH2023}
\usepackage{multirow}

\interspeechcameraready

\title{PCNN: A Lightweight Parallel Conformer Neural Network for Efficient Monaural Speech Enhancement}
\name{Xinmeng Xu$^{1}$, Weiping Tu$^{1,2,3,*}$ \thanks{$^*$ Corresponding Author.}, Yuhong Yang$^{1,3}$}
\address{
  $^1$NERCMS, School of Computer Science, Wuhan University,  China\\
  $^2$Hubei Luojia Laboratory, China\\
  $^3$Hubei Key Laboratory of Multimedia and Network Communication Engineering, Wuhan University, China}
\email{$\{$xuxinmeng, tuweiping, yangyuhong$\}$@whu.edu.cn}

\begin{document}

\maketitle
 
\begin{abstract}
Convolutional neural networks (CNN) and Transformer have wildly succeeded in multimedia applications. However, more effort needs to be made to harmonize these two architectures effectively to satisfy speech enhancement. This paper aims to unify these two architectures and presents a Parallel Conformer for speech enhancement. In particular, the CNN and the self-attention (SA) in the Transformer are fully exploited for local format patterns and global structure representations. Based on the small receptive field size of CNN and the high computational complexity of SA, we specially designed a multi-branch dilated convolution (MBDC) and a self-channel-time-frequency attention (Self-CTFA) module. MBDC contains three convolutional layers with different dilation rates for the feature from local to non-local processing. Experimental results show that our method performs better than state-of-the-art methods in most evaluation criteria while maintaining the lowest model parameters.
\end{abstract}

\section{Introduction}
Speech enhancement (SE) aims to estimate target speech from a noisy recording, which may consist of ambient noise, interfering speech, and room reverberation. It is a pre-processor for many speech processing applications, such as speech recognition \cite{ref1}, speaker verification \cite{ref2}, and hearing aids design \cite{ref3}. With the recent advances in supervised learning, deep neural networks (DNNs) are applied to several SE models. Typically, DNN-based SE models operate in the short-time Fourier transform (STFT) domain and estimate the clean target speech from the noisy signal via direct spectral mapping \cite{ref4,ref5}, or time-frequency (TF) masking \cite{ref6,ref7}.

Convolutional neural network (CNN) represents a successful backbone network architecture, which performs the filter processing on speech frames in parallel. It thus is structurally well-suited to focus on local patterns, such as harmonic structures \cite{ref8}. Meanwhile, CNN captures the contextual information by stacking multiple layers. While these properties bring efficiency and generalization to CNNs, they also cause two main issues. Firstly, the convolutional operation has a limited receptive field. Secondly, the convolution filters have static weights at inference. The former thus prevents the network from capturing the long-range feature dependencies \cite{ref9, ref10} while the latter sacrifices the adaptability to the input contents. As a result, it needs to meet the requirement in modeling the global noise distribution and generates results with noticeable noise residue. 

Self-attention (SA) calculates response at a given feature region by a weighted sum of all other positions \cite{ref11, ref12, ref13}. Benefiting from the advantage of global processing, SA achieves a significant performance boost over CNNs in SE tasks by mitigating their shortcomings, i.e., limited receptive field and inadaptability to input content \cite{ref14, ref15}. However, due to the global calculation of SA, its computation complexity grows quadratically with the spatial resolution, making it infeasible to fulfill the real-time demanding of SE systems. In addition, global relationships between these speech features are prone to bias and unreliable because feature regions are usually noisy \cite{ref5}. In this way, calculating the self-similarity of features between the target speech and the global mixture may not be a practical option.

\begin{figure*}[t]
  \centering
  \includegraphics[width=\linewidth]{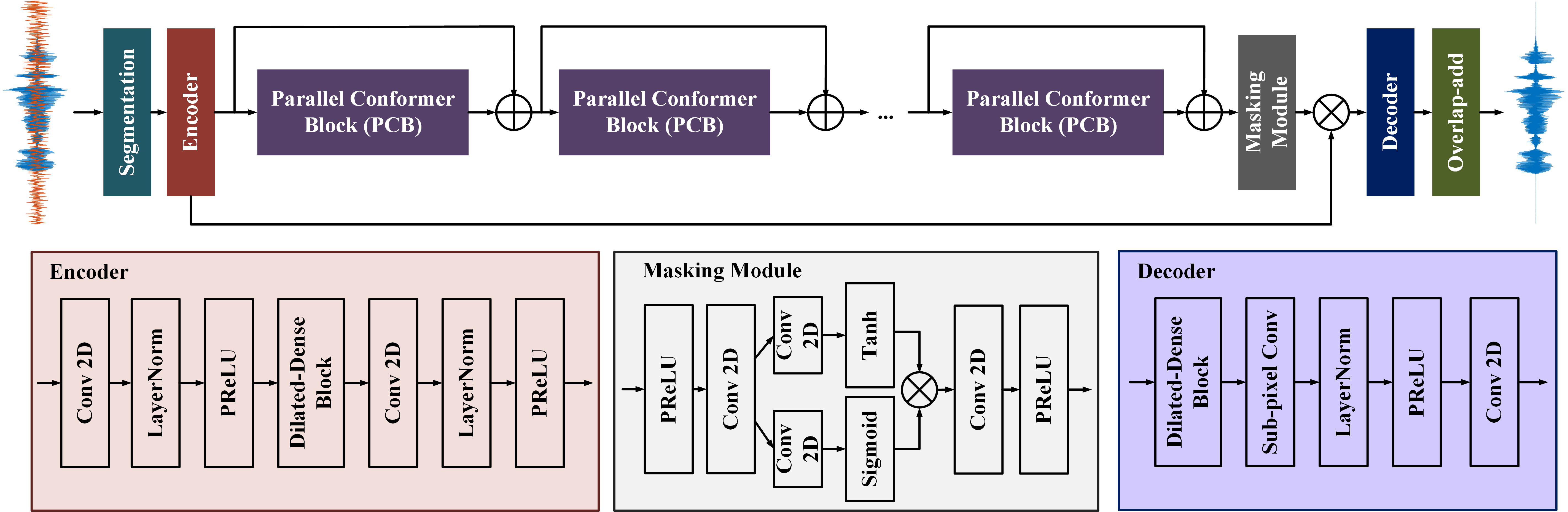}
  \caption{Overview of the parallel conformer neural network (PCNN).}
  \label{fig:1}
\end{figure*}

Inspired by the superior performance of CNN in extracting speech local format patterns and the effectiveness of Transformer in capturing the long-range dependency, we propose the parallel Conformer neural network (PCNN) for monaural speech enhancement. The proposed architecture incorporates CNN and Transformer in a parallel manner. It is followed by a hybrid fusion block containing depth-wise separable convolutions and channel attention for an adaptively and learnable performance trade-off. In addition, to deal with the small receptive field of CNN and the high computational complexity of the Transformer, we specially designed a multi-branch dilated convolution (MBDC) and a self-channel-time-frequency attention (Self-CTFA) module. In particular, the MBDC applies channel-wise attention to different dilation rates of convolutions to enlarge the size of the receptive field of local operation, in which channel-wise attention is independently performed on these three outputs for flexibly achieving feature processing from local to non-local. The self-CTFA module consists of three parallel attention branches, i.e., channel-dimension, time-dimension, and frequency-dimension, in which three 2D attention maps are calculated by three 1D energy distributions of these dimensions. 

\section{Model Description}
In this section, we elaborate on our proposed PCNN, which enables adaptive and learnable adjustment of contribution between CNN and Transformer in the SE tasks.

\subsection{Overview}
We propose a parallel conformer neural network (PCNN) for SE in the time domain. As shown in Figure~\ref{fig:1}, the architecture consists of a segmentation operation, encoder, separator, masking module, decoder, and overlap-add operation.

The input of PCNN is a raw speech waveform mixture, $\textbf{x}\in \mathbb{R}^{1\times N}$, which is firstly split into $F$ overlapped frames of length $L$ with a shifting size $S$ by \textbf{segmentation operation}. In this way, $\textbf{x}$ is resulted into a 3-dimensional tensor $X\in \mathbb{R}^{1\times F\times L}$, which is be expressed as
\begin{equation}
    F=\lceil(M-L)/(L-S)+1\rceil,
\end{equation}
where $N$ represents the length of the input speech mixture and $\lceil\cdot\rceil$ rounds the number involved up to the nearest integer. In addition, the \textbf{overlap-add operation} in an inverse of \textbf{segmentation operation} to merge the frames for the enhanced speech waveform reconstruction.

\textbf{Encoder} plays the role of feature extractor \cite{ref16, ref17} and contains two convolutional layers, of which the first one is increasing the number of channels to 64 using convolution with a kernel size of $(1,1)$ and the second one halves the dimension of frame size using a kernel size of $(1, 3)$ with a stride of $(1, 2)$, in which a dilated-dense block \cite{ref18} by using four dilated convolutions collaborates between them. In addition, layer normalization and PReLU \cite{ref19} are adopted after these convolutional layers. On the contrary, \textbf{decoder} is responsible for the feature reconstruction and contains a dilated-dense block, and a sub-pixel convolution \cite{ref20}, where followed by a layer normalization, PReLU, and a 2D convolutional layer with a kernel size of $(1, 1)$ for the channel dimension recovery of enhanced speech feature into 1.

\begin{figure}[t]
  \centering
  \includegraphics[width=0.6\linewidth]{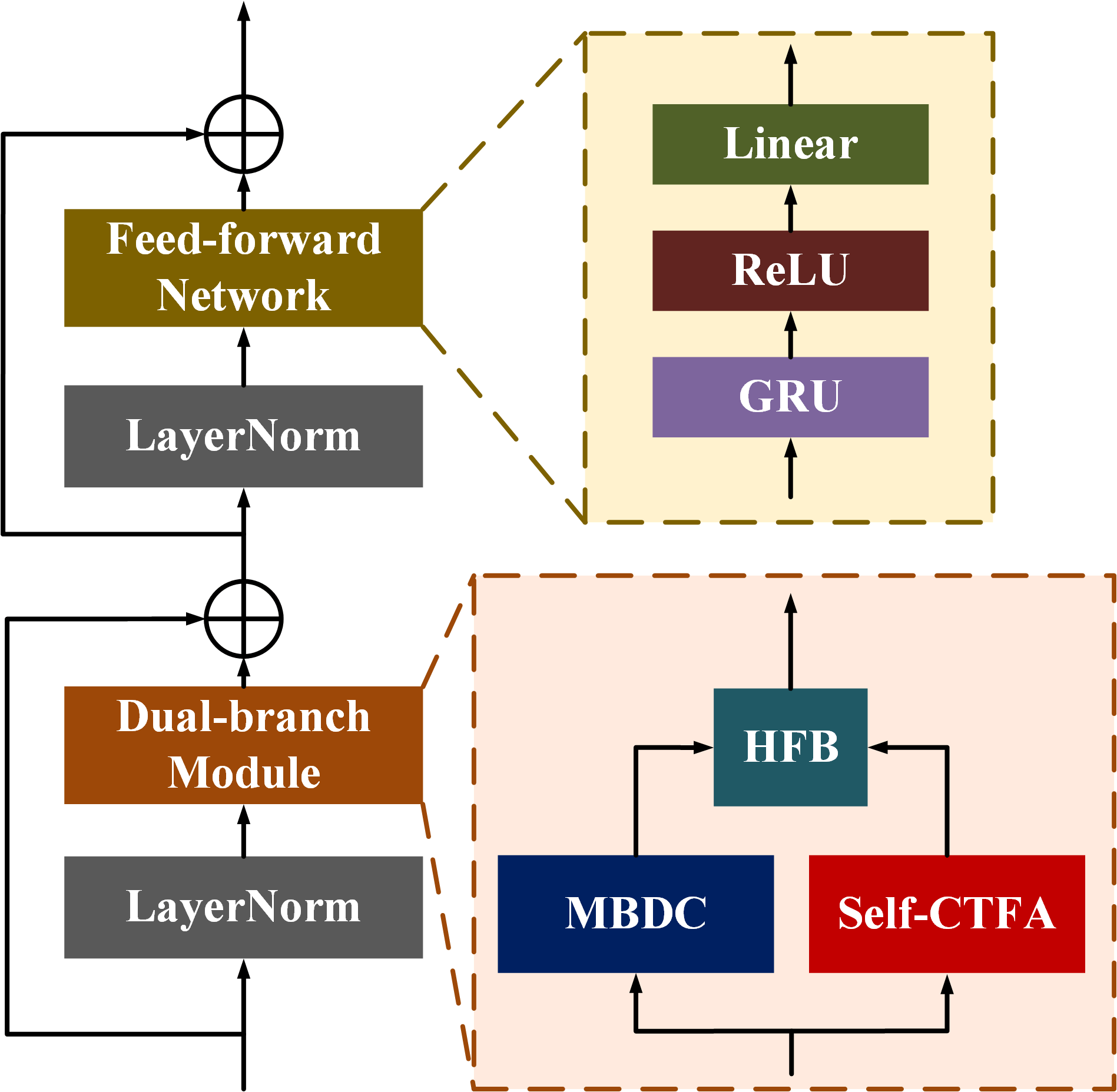}
  \caption{The architecture of parallel conformer block (PCB). ``MBDC'' denotes the multi-branch dilation convolution, ``Self-CTFA'' denotes self channel-time-frequency attention, and ``HFB'' denotes hybrid fusion block.}
  \label{fig:2}
\end{figure}

The \textbf{separator} is the crucial part of PCNN and is mainly composed of several parallel conformer blocks (PCBs) that are cascaded together. The PCB, as shown in Figure~\ref{fig:2}, include a dual-branch module containing an MBDC, Self-CTFA module, and HFB for local and global extracting and leveraging, and a feed-forward network, both preceded by layer normalization steps and with skip connections, which is specially described in Sec~\ref{sec:2.2}. Unlike conventional Transformer blocks, the feed-forward network consists of a gated recurrent unit (GRU) layer to learn the positional information \cite{ref21, ref22}. The \textbf{masking module} utilizes the feature output from \textbf{separator} to generate a mask to enhance speech. Concretely, the production from \textbf{separator} is doubled along the channel dimension with PReLU and convolution for matching the output of the encoder that then passes through a gated convolution operation \cite{ref23} and ReLU to get the mask. The element-wise multiplication between the mask and the output of the encoder obtains the final masked encoder feature.

Two loss functions are used in our study. One is the frequency-domain loss function, which starts with calculating the STFT to create a TF representation of the mixture sample. The  TF bins corresponding to the target speech are then separated and used to synthesize the source waveform using inverse STFT. In this case, the loss function is formulated by the mean square error (MSE) between the TF bins estimated target speech $\hat{S}\in \mathbb{R}^{T\times F}$ and the corresponding ground truth $S\in \mathbb\mathbb{R}^{T\times F}$,
\begin{equation}
    \mathcal{L}_f = \frac{1}{T\times F}||S - \hat{S}||^2
\end{equation}
where $||\cdot||^2$ denotes the $l_2$ norm, $T$, and $F$ denotes the number of frames and frequency bins, respectively.  We also use the time-domain loss based on the mean MSE between the enhanced speech and clean speech, which is defined as:
\begin{equation}
    \mathcal{L}_t = \frac{1}{N}\sum^{N}_{i=1}(x_i - \hat{x}_i),
\end{equation}
where $x$ and $\hat{x}$ are the clean speech and enhanced speech samples, respectively, and $N$ represents the number of samples. In this way, we obtain the final loss by combining these two types of loss functions, 
\begin{equation}
    \mathcal{L}_{total} = \alpha\mathcal{L}_f + (1-\alpha)\mathcal{L}_t,
\end{equation}
where $\alpha$ is a tunable parameter and is set as 0.2.
\begin{figure}[t]
  \centering
  \includegraphics[width=\linewidth]{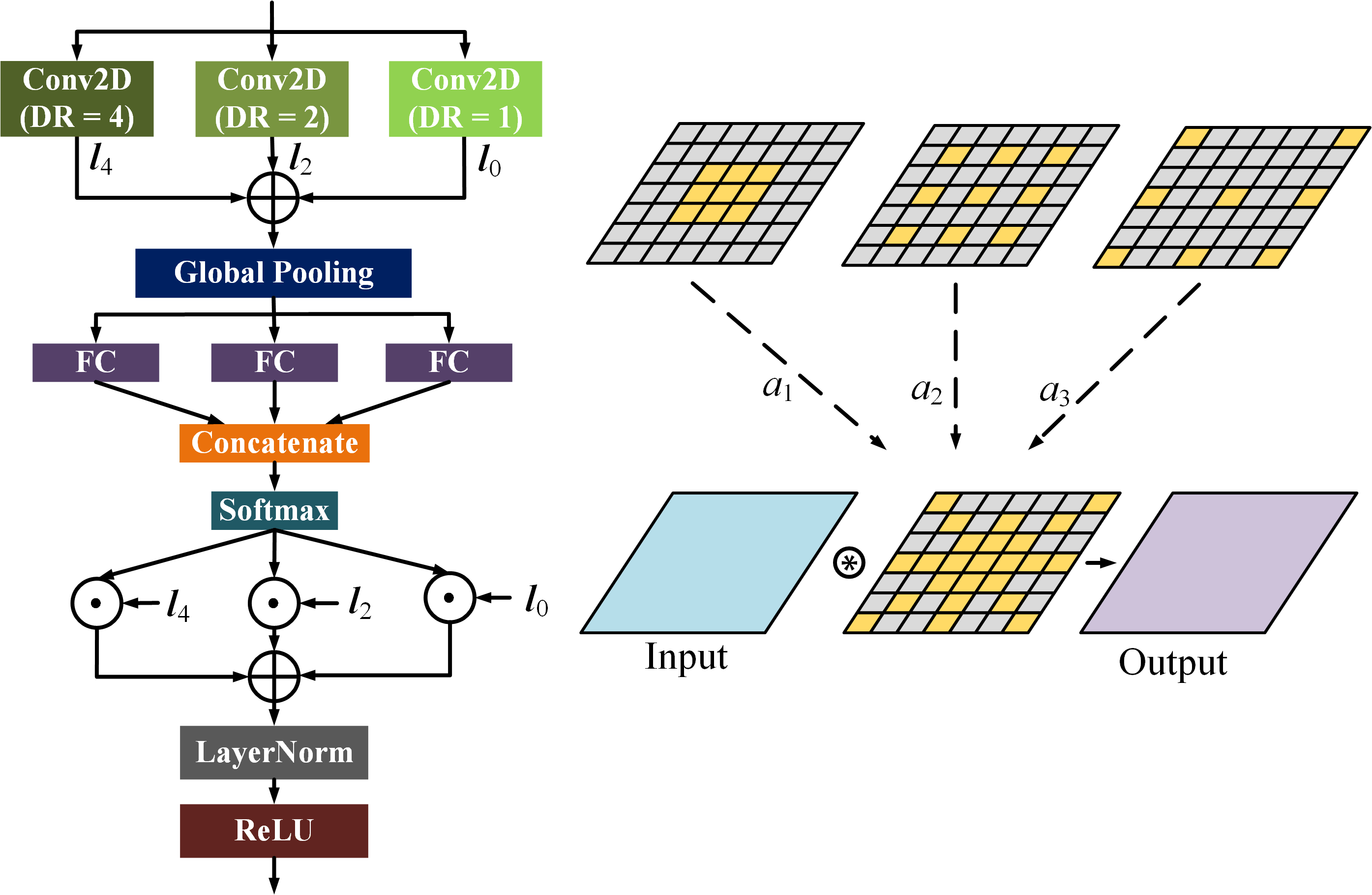}
  \caption{The architecture of Multi-branch Dilation Convolution (MBDC).}
  \label{fig:3}
\end{figure}

\subsection{Dual-branch Module}\label{sec:2.2}
Our proposed dual-branch module, labeled with an orange box in Figure~\ref{fig:2}, comprises three parts: MBDC for local processing, Self-CTFA module for global processing, and HFB for features fusion. The role of the dual-branch module is to leverage local and global operations adaptively.

\renewcommand{\arraystretch}{1.1}
\begin{table}[]
\centering
\caption{Comparison between MBDC, Dilated Convolution, and  Convolution.}
\begin{tabular}{l|ccc}
\hline
Type            & MBDC      & Dilated Conv    & Conv   \\ \hline
Kernel Size     & 3$\times$3 & 3$\times$3 & 9$\times$9 \\
Dilation Rate   & $\{$1, 2, 4$\}$ & 4         & 0        \\
Receptive Field & 9$\times$9  & 9$\times$9  & 9$\times$9 \\
Sampling Rate   & 33.33$\%$    & 11.11$\%$    & 100$\%$     \\
Param.(M)       & N         & N         & 9.00N    \\ \hline
\end{tabular}
\label{tab:1}
\end{table}

\textbf{Multi-branch Dilation Convolution (MBDC).} Inspired by \cite{ref24, ref25}, we employ the channel-wise attention mechanism to design the MBDC to perform channel selection with multiple convolutions with different dilation rates. The detailed architecture of our proposed MBDC is shown in Figure~\ref{fig:3}. In our design, we adopt three branches to carry different dilation rates of convolutional layers to generate feature maps with different receptive field sizes. The channel-wise attention is independently performed on these three outputs, and results are added together. In this way, the features can be extracted from local to non-local operations with the flexibility increasing, while the size of the receptive field is enlarged without substantial computational cost by the parallel structure of three dilated convolutional layers with the same kernel size \cite{ref26}. In addition, comparison results between MBDC, dilated convolution, and convolution in Table~\ref{tab:1} indicate that MBDC has a more significant feature sampling rate than dilated convolution while having much lower computational complexity than conventional convolution.

\begin{figure}[t]
  \centering
  \includegraphics[width=0.95\linewidth]{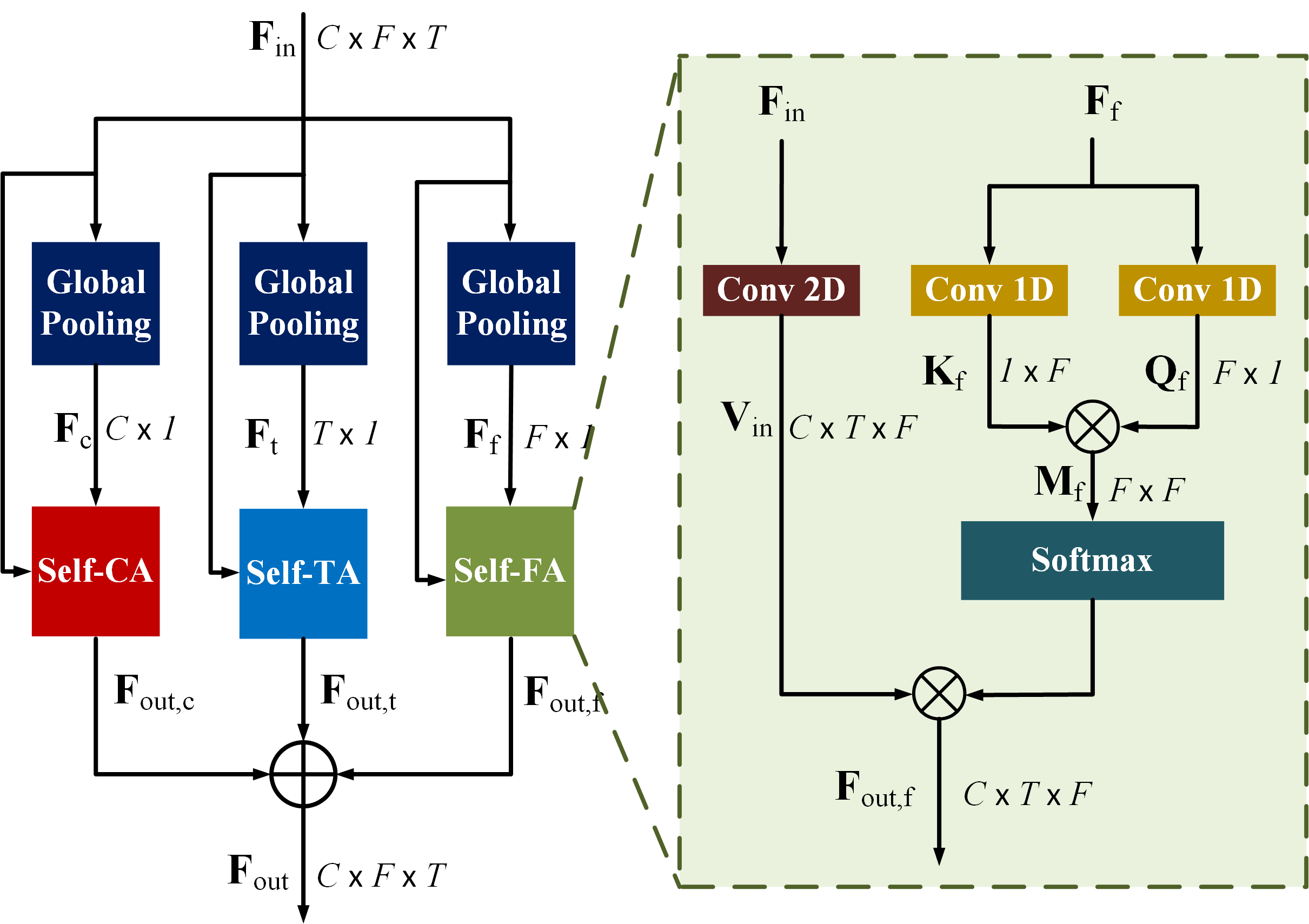}
  \caption{The architecture of Self Channel-Time-Frequency Attention (Self-CTFA) Module.}
  \label{fig:4}
\end{figure}

\textbf{Self Channel-Time-Frequency Attention (Self-CTFA) Module.} We show the proposed Self-CTFA module in Figure~\ref{fig:4}. The Self-CTFA module takes TF representation $\textbf{F}_{in}\in \mathbb{R}^{C\times T\times F}$ as input, where $C$, $T$, and $F$ denote the channels, time frames, and frequency bins, respectively. Self-CTFA module consists of three branches to generate a 1D channel-dimension energy feature $\textbf{F}_{c}\in \mathbb{R}^{C\times 1}$, a 1D time-dimension energy feature $\textbf{F}_{t}\in \mathbb{R}^{1\times T}$, and a 1D frequency-dimension energy feature $\textbf{F}_{f}\in \mathbb{R}^{F\times 1}$ in parallel by separately using three global pooling functions. 

Each branch contains two sub-branches to calculate query and key using two 1D $1\times 1$ convolutional layers. After that, the query and key of each  branch are multiplied and fed into the softmax activation function to generate the attention feature map, which is defined as:
\begin{equation}
\left\{
    \begin{array}{lr}
        \textbf{M}_c = \text{softmax}(\mathcal{H}_{c1}(\textbf{F}_c) \times \mathcal{H}_{c1}(\textbf{F}_c)^{\top}),\\
       \textbf{M}_t = \text{softmax}(\mathcal{H}_{c1}(\textbf{F}_t) \times \mathcal{H}_{c1}(\textbf{F}_t)^{\top}),\\
       \textbf{M}_f = \text{softmax}(\mathcal{H}_{c1}(\textbf{F}_f) \times \mathcal{H}_{c1}(\textbf{F}_f)^{\top}),
    \end{array}
\right.
\end{equation}
where $\textbf{M}_c\in \mathbb{R}^{C\times C}$, $\textbf{M}_t\in \mathbb{R}^{T\times T}$, and $\textbf{M}_f\in \mathbb{R}^{F\times F}$ denote the attention feature maps of channel branch, time branch, and frequency branch, respectively, and $\mathcal{H}_{c1}$ represents the 1D $1\times 1$ convolutional layer. Afterwards, the $\textbf{V}_{in}$ generated from $\textbf{F}_{in}$ separately multiples with $\textbf{M}_c$, $\textbf{M}_t$, $\textbf{M}_f$, and these results are added to obtain the output of Self-CTFA module $\textbf{F}_{out}$. In this way, the Self-CTFA module reduces the computational complexity from $\mathcal{O}(C^2TF+CT^2F+CTF^2)$ to $\mathcal{O}(C^2+T^2+F^2)$.

\textbf{Hybrid Fusion Module (HFB).} Considering the feature redundancy and knowledge discrepancy among MBDC and Self-CTFA module, we introduce a novel hybrid fusion block (HFB) in our approach.  Specifically, we incorporate depth-wise separable convolutions and the channel attention layer into HFB to discriminatively aggregate features in spatial and channel dimensions \cite{ref26.5}.

\renewcommand{\arraystretch}{0.95}
\begin{table*}[t]
\centering
\caption{Comparisons of baseline models in terms of STOI, PESQ, and SSNR.}
\resizebox{\textwidth}{!}{
\begin{tabular}{l|ccc|ccc|ccc|cc}
\hline
Test SNR            & \multicolumn{3}{c|}{-5 dB} & \multicolumn{3}{c|}{0 dB} & \multicolumn{3}{c|}{5 dB} & \multirow{2}{*}{Param.} & \multirow{2}{*}{RTF} \\ \cline{1-10}
Metric              & STOI (\%)    & PESQ   & SSNR    & STOI (\%)   & PESQ   & SSNR   & STOI (\%)   & PESQ   & SSNR   &                         &                      \\ \hline
Unprocess           & 57.71   & 1.37   & -5.04   & 71.02   & 1.73   & -0.05  & 82.53   & 2.03   & 4.93   & -                       & -                    \\ \hline
Conv-TasNet         & 79.81   & 2.23   & 6.65    & 86.76   & 2.42   & 8.23   & 91.46   & 2.86   & 10.12  & 4.58 M                  & 0.72                 \\
GRN                 & 80.31   & 2.21   & 6.62    & 86.98   & 2.49   & 8.31   & 91.28   & 2.91   & 10.89  & 3.08 M                  & 0.68                 \\
TSTNN               & 83.76   & 2.32   & 6.98    & 89.75   & 2.64   & 8.86   & 93.67   & 3.03   & 11.59  & 4.86 M                  & 0.86                 \\
U-Former            & 84.75   & 2.38   & 7.03    & 89.60   & 2.68   & 8.94   & 93.76   & 3.08   & 11.96  & 5.85 M                  & 0.88                 \\ \hline
PCNN                & 87.24   & 2.51   & 7.84    & 92.17   & 2.83   & 9.71   & 94.82   & 3.13   & 12.34  & 3.15 M                  & 0.51                 \\
PCNN (DC)           & 85.69   & 2.44   & 7.11    & 90.64   & 2.71   & 9.38   & 93.92   & 3.09   & 12.01  & 3.13 M                  & 0.49                 \\
PCNN (CC)           & 85.98   & 2.46   & 7.69    & 91.86   & 2.74   & 9.39   & 94.19   & 3.11   & 12.14  & 3.57 M                  & 0.54                 \\
PCNN (SA)           & 87.36   & 2.50   & 7.84    & 92.06   & 2.83   & 9.75   & 94.79   & 3.12   & 12.28  & 6.36 M                  & 0.96                 \\
PCNN -w/o MBDC      & 83.32   & 2.30   & 6.86    & 87.96   & 2.59   & 8.78   & 92.89   & 2.98   & 11.87  & 2.98 M                  & 0.46                 \\
PCNN -w/o Self-CTFA & 82.67   & 2.26   & 6.79    & 87.01   & 2.51   & 8.66   & 91.47   & 2.93   & 10.98  & 2.74 M                  & 0.43                 \\ \hline
\end{tabular}}
\label{tab:2}
\end{table*}

\section{Experimental Setup}

\subsection{Datasets}

In order to evaluate the performance of the proposed model, experiments are conducted on Librispeech corpus \cite{ref27}. There 6500 clean utterances are selected for the training set and 400 for the validation set, created under the random SNR levels ranging from -5dB to 10 dB. The test set contains 100 utterances under the SNR condition of -5dB, 0dB, 5dB, and 10 dB.

Noise signals from the Demand dataset \cite{ref28}, along with the clean speech recordings, are used to create the noisy speech for the training and validation set.  The clean speech and noise recordings with a sampling frequency of 16 kHz and the frame size and frameshift for frame-level processing are set to 512 and 256.

\subsection{Training and Network Parameters}
The clean speech and noise recordings with a sampling frequency of 16 kHz and the frame size and frameshift for frame-level processing are set to 512 and 256. In each training epoch, we chunk a random segment of 4 seconds from an utterance if it is more significant than 4 seconds. The smaller utterances are zero-padded to match the size of the largest utterance in the batch. The Adam optimizer is used for stochastic gradient descent (SGD) based optimization, and the initial learning rate is set to 0.001. MSE is used as a loss function.

\begin{table}[]
\centering
\caption{Ablation study of self-CTFA by removing different components in -5 dB SNR condition.}
\begin{tabular}{l|ccc}
\hline
Metric                       & STOI (\%) & PESQ & SSNR \\ \hline
PCNN                  & 87.24     & 2.51 & 7.84 \\ \hline
\textit{-w/o C Branch}       & 84.81     & 2.47 & 7.51 \\
\textit{-w/o T Branch}       & 85.99     & 2.44 & 7.46 \\
\textit{-w/o F Branch}       & 85.56     & 2.41 & 7.49 \\
\textit{-w/o C-T Branches}   & 84.57     & 2.35 & 7.23 \\
\textit{-w/o C-F Branches}   & 84.59     & 2.32 & 7.15 \\
\textit{-w/o T-F Branches}   & 83.98     & 2.30 & 7.13 \\
\textit{-w/o C-T-F Branches} & 83.05     & 2.26 & 6.98 \\ \hline
\end{tabular}
\label{tab:3}
\end{table}

\section{Results and Analysis}
\subsection{Model Comparison}
This section compares alternative baseline models in Table~\ref{tab:2} in terms of STOI, PESQ, and SSNR, where the numbers represent the averages over the test set in each condition. Four baseline systems are selected for the comparison, i.e., Conv-TasNet \cite{ref17}, GCRN \cite{ref23}, TSTNN \cite{ref29}, and U-Former \cite{ref14}. In addition, we also evaluate the performance of PCNN when replacing MBDC with dilated convolution, i.e., PCNN (DC), replacing MBDC with conventional convolution, i.e., PCNN (CC), replacing self-CTFA module with self-attention module, i.e., PCNN (SA), removing MBDCs, i.e., PCNN -w/o MBDC, and removing self-CTFA module, i.e., PCNN -w/o Self-CTFA.

Table~\ref{tab:2} shows the comparison results of the proposed PCNN and the other four baseline models. The number of parameters and real-time factory (RTF) is also presented. One can observe the following phenomena. First, the proposed model consistently outperforms all baselines in three metrics scores for different cases. Secondly,  the proposed PCNN has the fewest parameters followed by GRN and lowest RTF, but PCNN has the best performance, demonstrating the higher performance effectiveness of PCNN. Thirdly, we compare the PCNN when using different components to replace MBDC and self-CTFA module in Table~\ref{tab:2}, according to the results, we observe that (1) replacing MBDC with dilated convolution achieves lower scores in these evaluation metrics with a similar number of parameters, (2) replacing MBDC with conventional convolution that has the exact size of the receptive field with dilated convolution achieves similar performance with a higher number of parameters and RTF. Finally, the comparison results between PCNN, PCNN -w/o MBDC, and PCNN -w/o Self-CTFA indicate the necessity of the proposed MBDC and self-CTFA module.

\subsection{Impact of Branches in Self-CTFA Module}
The proposed self-CTFA module consists of the channel (C) branch, time (T) branch, and frequency (F) branch. In this study, we evaluate variants of the self-CTFA module when removing the C, T, and F branches. We set the same parameters for the algorithm in the previous section but control the usage of different branches. Table~\ref{tab:2} demonstrates the evaluation results of the self-CTFA module when removing other components in the -5 SNR condition. According to Table~\ref{tab:2}, we conclude that the existence of the C branch, T branch, and F branch does promote the performance of self-CTFA module. In addition, the F branch performs better than the T and C branches.

\section{Conclusion}
This paper proposes a parallel conformer neural network (PCNN) for SE by leveraging CNN for local detail information capture and the transformer for long-range dependencies extraction. According to the drawbacks of CNN and SA in the transformer, we develop an MBDC to address the small receptive field of CNN and a self-CTFA model to address the high computational complexity of SA. In addition, an HFB is utilized for leveraging the MBDC and self-CTFA. Through experiments, we show the superiority of the proposed method over other methods compared in this paper.

\section{Acknowledgements}
This work is supported by the National Nature Science Foundation of China (No. 62071342, No.62171326), the Special Fund of Hubei Luojia Laboratory (No. 220100019), the Hubei Province Technological Innovation Major Project (No. 2021BAA034) and the Fundamental Research Funds for the Central Universities (No.2042023kf1033).

\vfill\pagebreak
\bibliographystyle{IEEEtran}
\bibliography{mybib}

\end{document}